\documentclass[fleqn,usenatbib,useAMS]{mnras}

\usepackage[T1]{fontenc}
\usepackage{ae,aecompl}

\usepackage{url}
\vspace*{\fill}

\usepackage{graphicx}	
\usepackage{natbib}
\usepackage[switch]{lineno}
\defcitealias{Gir2016}{G16}

\usepackage[switch]{lineno}
						  \usepackage{multirow} 
						  \usepackage{cancel}   
						  \usepackage{adjustbox}

\usepackage{atbegshi}
\AtBeginDocument{\AtBeginShipoutNext{\AtBeginShipoutDiscard}}

\title{Radio spectral properties of cores and extended regions in blazars in the MHz regime}

\author[D. d'Antonio et al.]{
D. d'Antonio,$^{1}$\thanks{E-mail: daniele.dantonio364@gmail.com}
M. Giroletti,$^{2}$
G. Giovannini$^{1,2}$
and A. Maini$^{1,2,3}$
\\
$^{1}$Dipartimento di Fisica e Astronomia, Universit\`a di Bologna, via Gobetti 93/2, 40129 Bologna, Italy\\
$^{2}$INAF -- IRA, via Gobetti 101, 40129 Bologna, Italy\\
$^{3}$Department of Physics and Astronomy, Macquarie University, Balaclava Road, North Ryde, NSW, 2109, Australia
}

\date{Accepted XXX. Received YYY; in original form ZZZ}

\pubyear{2019}

\begin{document}
\label{firstpage}
\pagerange{\pageref{firstpage}--\pageref{lastpage}}
\maketitle

\begin{abstract}
Low-frequency radio surveys allow in-depth studies and new analyses of classes of sources previously known and characterised only in other bands. In recent years, low radio frequency observations of blazars have been available thanks to new surveys, such as the GaLactic and Extragalactic All-sky MWA Survey (GLEAM). We search for gamma-ray blazars in a low frequency ($\nu<240$\,MHz) survey, to characterise the spectral properties of the spatial components. We cross-correlate GLEAM with the fourth catalogue of active galactic nuclei (4LAC) detected by the \textit{Fermi} satellite. This improves over previous works using a low frequency catalogue that is wider, deeper, with a better spectral coverage and the latest and most sensitive gamma-ray source list.In comparison to the previous study based on the commissioning survey, the detection rate increased from 35\% to 70\%. We include Australia Telescope 20\,GHz (AT20G) Survey data to extract high-frequency high-angular resolution information on the radio cores of blazars. We find low radio frequency counterparts for 1274 out of 1827 blazars in the 72-231 MHz range. Blazars have flat spectrum at $\sim$ 100\,MHz regime, with a mean spectral index $\alpha = -0.44 \pm 0.01$  (assuming $S_{\nu} \propto \nu^{\alpha}$). Low synchrotron peaked objects have a flatter spectrum than high synchrotron peaked objects. Low frequency radio and gamma-ray emission show a significant but scattered correlation. The ratio between lobe and core radio emission in gamma-ray blazars is smaller than previously estimated.
\end{abstract}




\section{Introduction}

Blazars are the most extreme objects in the class of AGN (Active Galactic Nuclei). In these objects, the relativistic plasma jets produced by the central supermassive black hole are closely aligned with the line of sight, which result in an emission amplified by the Doppler boosting effect. For this reason, these objects are detected at cosmological distances. Their emission is mainly non-thermal, with a spectral energy distribution (SED) characterised by two humps (\citealt{Bla1978}, \citealt{Abdo2010I}, \citealt{Urry&Padovani1995}).

Blazars are composed by two main classes: BL Lac objects (BL Lacs) with weak emission lines in optical spectra, and flat-spectrum radio quasars (FSRQ) with strong emission lines in optical band (\citealt{sti1991}). Furthermore, blazars can be classified based on the position of the synchrotron peak frequency. Low synchrotron peaked (LSP) sources have peak frequency smaller than \mbox{$\nu_{\mathrm{peak}}$ < 10$^{14}$\,Hz}; intermediate synchrotron peaked (ISP) sources have peak frequency in the range \mbox{10$^{14}$\,Hz < $\nu_{\mathrm{peak}}$ < 10$^{15}$\,Hz}; and high synchrotron peaked (HSP) sources have peak frequency greater than \mbox{$\nu_{\mathrm{peak}}$ > 10$^{15}$\,Hz} (\citealt{Ack2015}).

In this work we present the results achieved on the investigation of the spectral properties of gamma-ray blazars carried out thanks to the new, wide-area, high-sensitivity, large-bandwidth sky survey at low radio frequencies: the GaLactic and Extragalactic All-sky MWA Survey (GLEAM, \citealt{HW2017}), produced by the Murchison Widefield Array (MWA, \citealt{Tin2013}). GLEAM covers about 24\,831 sq.\ deg in the southern sky with a bandwidth between 72\,MHz and 231\,MHz, and contains 307\,455 objects. The advantage of GLEAM is that it allows for the first time to perform an accurate study of blazars at low radio frequency in the MHz regime.  

Thanks to the advent of the Large Area Telescope (LAT) on board of the \textit{Fermi} satellite, moreover, in the last few years the astrophysics of gamma-rays has greatly improved. We cross-correlate the GLEAM catalogue with the fourth catalogue of active galactic nuclei detected by the \textit{Fermi} satellite (4LAC, \citealt{4LAC}), to study the connection between the radio band and the gamma band. The 4LAC catalogue is a sub-sample of gamma-ray blazars, extracted by the LAT collaboration from the the fourth \textit{Fermi}-LAT catalogue of gamma-ray sources (4FGL, \citealt{4FGL}), which is the latest and deepest  source list in the energy range from 50 MeV to 1 TeV.

In recent years, several studies carried out at low frequency (i.e., below $\sim$1GHz)nhave allowed us to learn more information about blazars. The discovery that blazars have flat radio spectra in the regime of hundred of MHz is particularly intriguing. Indeed, flat radio spectra are expected at GHz regime \citep{Healey2007,Ivezi2002,Kimball2008}, but only recently it has become evident that this flat spectral behaviour is still present at lower frequencies. \citet{Mas2013,Mas2014} found gamma-ray blazars with radio counterparts which show flat spectra in the interval between $\sim$325 MHz and 1.4 GHz. This result was confirmed by \citet{Nor2014} in a slightly less wide frequency range (352-1400 MHz). \citet{Mas2013I} performed a study on a wider frequency interval (74-1400 MHz and 1.4-4.85 GHz), for which 60\% of gamma-ray blazars have flat spectra and 99\% of the whole sample present radio spectra evidently influenced by a flat core component. The work suggests that blazar spectra are characterised by beaming effect due to particles accelerated in relativistic jets even down to 74 MHz. Furthermore, \citet{Gir2016}, \citet{Fan2018} and \citet{Mooney2019} highlighted a gamma/radio correlation which, however, becomes weaker at lower radio frequency. All these studies show that the flat core component has an important contribution at low radio frequencies (from 74 to 1400 MHz).

The previous work \citep[][hereafter \citetalias{Gir2016}]{Gir2016} cross-correlated the 3LAC catalogue with the Murchison Widefield Array Commissioning Survey Compact Low-Frequency Source (MWACS, \citealt{HW2014}), a preliminary survey carried out with the MWA. Given the wider sky area, the deeper sensitivity, and the better spectral coverage of GLEAM with respect to MWACS, we decided to significantly improve the constraints derived in the previous work about the low frequency properties (hundred of MHz) of gamma-ray blazars. We find 1274 out of 1827 gamma-ray blazars. This number of sources with a radio counterpart that we analysed is larger than in another work based on LOFAR  \citep{Mooney2019} by a factor of 13. The number of objects in our sample is also larger than GMRT \citep{Fan2018} data by a factor of 1.3. In detail, \citet{Mooney2019} have a total number of 102 sources and find radio counterpart to all the 98 objects from the Third Fermi-LAT Source Catalogue (3FGL, \citealt{Ace2015}). Instead, \citet{Fan2018} find 983 out of 1328 sources from the TIFR GMRT Sky Survey \citep[TGSS,][]{Int2017}. Moreover, we use the Australia Telescope 20\,GHz survey (AT20G, \citealt{Mur2010}) to characterise the nuclear emission, when possible.

In Sect.\,\ref{sec:Catalogues} all the catalogues are described, including properties such as their frequency range, sky area, number of sources and the counts and types of objects for every survey. In Sect.\,\ref{sec:Cross-corr_catalogue} we explain the selection criteria of the samples we utilised (gamma-ray blazars in the GLEAM and AT20G surveys). The results about the spectral properties are presented in Sect.\,\ref{sec:Results}, while in Sect.\,\ref{sec:Conclusions} we report the final discussion.

In this paper, we adopt a $\Lambda$CDM cosmology with \mbox{$h$ = 0.71}, $\Omega_{m}$ = 0.27, and $\Omega_{\Lambda}$ = 0.73 \citep{Kom2009}. Also, the radio spectral index $\alpha$ is defined by the flux density $S_{\nu}$ and the frequency $\nu$ adopting the convection $S_{\nu} \propto \nu^{\alpha}$, while the gamma-ray photon index $\Gamma$ respects the relation $dN_{\mathrm{photon}}/dE \propto E^{-\Gamma}$ where it is defined by the photon flux $dN$ i.e. the number of photons per square centimetre per second, and the energy E (note that \citetalias{Gir2016} used $S_{\nu} \propto \nu^{-\alpha}$ and $dN_{\mathrm{photon}}/dE \propto E^{-\Gamma}$).

\section{Catalogues}
\label{sec:Catalogues}

\subsection{GLEAM catalogue}
\label{subsec:GLEAM_catalogue}

GLEAM observations began in August 2013. The first two runs of observations (concluded in July 2014 and July 2015, respectively), where carried out in the frequency range $\sim$72--231\,MHz. A third run of observations, concluded in July 2016, was performed at higher radio frequencies ($\sim$250--310\,MHz), which are not part of this work as we were interested in analysing the lowest frequency range. The GLEAM survey covers the entire south sky up to $+30^{\circ}$ of celestial declination, excluding the Galactic latitudes within 10\,deg from the Galactic plane and the Magellanic Clouds \citep{HW2017}. This is also the sky area where GLEAM overlaps the Australia Telescope 20\,GHz Survey (AT20G). The final catalogue is mainly composed of extra-galactic sources.  

GLEAM has a resolution down to \mbox{$\sim$2\,arcmin}, and in total contains 307\,455 radio sources. For every source both peak and integrated flux density centred at 200\,MHz are reported, as well as 20 separate flux density measurements across the 72--231\,MHz range, each one based on a bandwidth 7.68\,MHz wide. The survey is reported to be 90\% complete at 170\,mJy, and 50\% complete at 55\,mJy (see \citealt{HW2017} for more details).

\subsection{AT20G catalogue}
\label{subsec:AT20G_catalogue}

The Australia Telescope 20\,GHz Survey (AT20G) is a radio survey performed in the period between 2004 and 2008, at 20\,GHz with the Australia Telescope Compact Array (ATCA). It covers the whole sky south of 0$^\circ$ degrees declination. Therefore, the AT20G covers the whole south sky area which is covered by GLEAM. It is the largest and most sensitive survey of the sky every carried out at this frequency, with a flux density limit above 40\,mJy.

The final catalogue is composed by 5890 objects, all the sources with a flux density greater than 50\,mJy at 20\,GHz. Moreover, some sources are also provided of 5 and 8\,GHz flux densities. This catalogue has been fundamental for the study of radio emission in flat-spectrum sources, such as the cores of radio-loud AGNs \citep{Mah2010}. We used the data from version 1.0 in Vizier.

\subsection{4FGL catalogue}
\label{subsec:4FGL_catalogue}

The fourth catalogue of the \textit{Fermi}-LAT satellite (4FGL; \citealt{4FGL}) is a catalogue of gamma-ray sources observed by \textit{Fermi} over the first 8 years of the mission, based on data in the energy range from 50 MeV to \mbox{1 TeV}. With twice as much exposure as well as a number of analysis improvements, including an updated model for the Galactic diffuse gamma-ray emission, it supersedes the third \textit{Fermi}-LAT catalogue (3FGL) and the earlier releases: the LAT Bright Source List \citep[0FGL,][]{0FGL}, the First \citep[1FGL,][]{1FGL} and the Second \textit{Fermi}-LAT \citep[2FGL,][]{2FGL} catalogues. Moreover, including an update model for the Galactic diffuse gamma-ray emission, the 4FGL catalogue supersedes also its early release (FLY8\footnote{\url{https://fermi.gsfc.nasa.gov/ssc/data/access/lat/fl8y/}}). The 4FGL catalogue is deeper than the previous catalogues in the 0.1--300\,GeV energy range.

The total number of sources in the 4FGL is 5065 above $4\sigma$, for each of which are reported position, significance, photon index and gamma-ray flux. The catalogue also lists the flux in various energy ranges, as well as the total flux in yearly and bi-monthly time intervals constituting the 8 years of observations. The distribution of the 95\% confidence error radius (geometric mean of semi-major and semi-minor axes of the 95\% confidence error ellipse) of the sample has the highest peak at $\sim 4$ arcmin. This is verified for all sources at high Galactic latitudes ($|b| > 10^\circ$). Most sources are associated with blazars, while a significant fraction ($\sim 26\%$) does not have a counterpart at low frequencies. We accessed the complete catalogue on the \textit{Fermi} Science Support Center web site\footnote{\url{https://fermi.gsfc.nasa.gov/ssc/data/access/}}.

\subsubsection{4LAC catalogue: a subset of the 4FGL}
\label{subsec:4LAC_catalogue:_a_subset_of_the_4FGL}

The fourth LAT AGN catalogue (4LAC) is a subset of the 4FGL catalogue that includes all gamma-ray sources at high Galactic latitudes ($|b| > 10^\circ$) associated with AGNs, either as a result of a Bayesian association \citep{Abd2010} or based on the likelihood ratio technique \citep{Ackermann2011b}. At the moment of writing, the publicly available version of the 4LAC\footnote{\url{ftp://www.cenbg.in2p3.fr/astropart/Fermi/4LAC/}} includes 2863 blazars.  We further restrict this list to the 2682 objects for which we were able to recover the radio flux density from the NVSS or the SUMSS.

Because the 4LAC is a gamma-ray catalogue, the most abundant type of sources are blazars (98\% of the total sources, both as confirmed and candidates). The catalogue contains also information about the low frequency counterparts (as the VLBI identifier and position, and the {\it Gaia} identifier), the redshift, and the position of the peak of the synchrotron component of the SED. According to the SED classification, the subset is composed of 1125 LSPs, 399 ISPs, and 362 HSPs. According to the spectroscopic classification, the subset is composed of 644 FSRQs, 1036 BL Lacs, and 939 blazar candidates of unknown/uncertain type (BCU). This last category includes both confirmed blazars of uncertain type (i.e., included in the BZCat catalogue --\citealt{Mas2015}-- and provided with the optical spectrum) and blazar candidates lacking of a proper optical spectroscopic analysis but characterised by the typical properties of blazars (e.g., two humped SED or flat radio spectrum). However, several optical spectroscopic investigations such as, \citet{Massaro2016opt}, \citet{Marchesi2018}, \citet{Paino2017I}, \citet{Paiano2017II}, \citet{Pena-Herazo2017} and \citet{Alvarez2016}  have demonstrated that most BCUs are evidently BL Lacs.

\section{Cross-correlation of the catalogues}
\label{sec:Cross-corr_catalogue}

The cross-matching of the catalogues was performed using TOPCAT \citep{Tay2005}. In this section we provide some details on the procedure we applied.

\begin{figure}
	\hspace{-0.2cm}
	\includegraphics[scale=0.9]{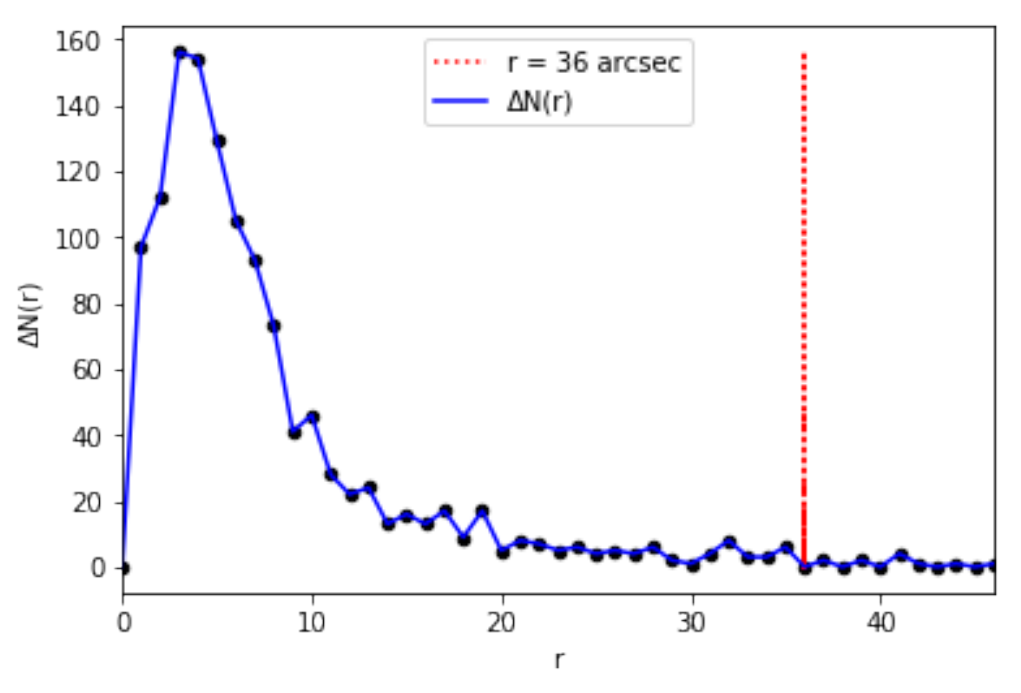}
	\caption{\label{fig:conteggi} Difference $\Delta N(r)$ between the number of cross-matches $N(r)$ at a given radius r and the same quantity at $r-\Delta r$ between 0'' and 46''. Note also that $\Delta r= 1''$. The red dotted line highlights the matching radius at 36\arcsec\  which is the value we have adopted to create the sample 4LAC-GLEAM.}
\end{figure}

\subsection{Cross-match 4LAC-GLEAM}
\label{subsec:Cross-match_4LAC-GLEAM}

Initially, we followed the same procedure that we had adopted for the 3LAC-MWACS cross match in \citetalias{Gir2016}: for the blazar positions, we assumed a 5\arcsec\ error radius, since the radio counterparts of LAT blazars generally come from interferometric surveys in the GHz domain, such as AT20G, NRAO VLA Sky Survey (NVSS, \citealt{Con1998}), Sydney University Molonglo Sky Survey (SUMSS, \citealt{Mau2003}), Faint Images of the Radio Sky at Twenty-centimeters (FIRST, \citealt{Hel2015}), and Australia Telescope PMN follow-up survey (PMN, \citealt{McC2012}); they often even have sub-milliarcsecond accuracy as provided by VLBI observations; for the GLEAM sources we considered the 95\% error ellipse obtained as $1.621\times$ the uncertainties reported on the RA and Dec coordinates.  This resulted in a total of 1012 matches, which was already a significant increase over the 3LAC-MWACS results, both in terms of sheer number of sources (by a factor $5.4\times$) and detection rate (with a 15\% increase).   
To assess the number of spurious matches we created 100 fake gamma-ray catalogues by shifting the position of the blazars by 2 degrees in random directions, and cross correlated these random lists with GLEAM.   Most of these cross-comparisons resulted in none or very few match (1.55 on average) so that the fraction of spurious matches in our list turns out to be very small (0.0015).   Since this fraction remains small even increasing the matching radii, while the number of real matches keeps increasing, we concluded that the use of the nominal positional uncertainties was actually leading us to underestimate the number of real matches.  One possible reason is that the centroid of the MHz and GHz emission regions are offset from each other by more than their respective positional uncertainties. We might expect a weak contribution from the extended region. Consequently, even a slight misalignment of the jet axis (e.g. $10^\circ$), for a 100 kpc lobe at $z=0.05$ can result in a projected displacement of $\sim 17\arcsec$ from the core, and thus result in a missing match.

Therefore, we adopted the procedure developed by \citet{Mas2013,Mas2013I}, and determined a matching radius of 36\arcsec, above which the number of matches obtained using the simulated gamma-ray catalogues exceeds that of the real catalogue. Following the same approach used by \citet{Mas2013I}, Fig. \ref{fig:conteggi}  highlights the difference between the number of cross-matches $N(r)$ at a given radius $r$ and the same quantity at $r -\Delta r$: $ \Delta N(r) = N(r) - N(r -\Delta r)$, where $\Delta r = 1''$.

Our cross-matching provided us with 1274 objects, all unique. Using the same technique described above (simulating 100 gamma-ray catalogues, each one obtained by shifting every original gamma-ray blazar by $2^\circ$ in a random direction), we estimate the fraction of spurious associations in our final list to be $\sim 0.0055$.

According to the SED classification, the matched sources are  716 LSPs, 130 ISPs, and 115 HSPs; while according to the spectroscopic classification these matched sources are 406 FSRQs, 384 BL Lacs, and 456 BCUs. The value of 1274 matches corresponds to a high detection rate of $\sim 70\%$, implying that most blazars are detected at low radio frequency with the GLEAM sensitivity.  To check that all matches are unique, we both searched for the GLEAM counterparts of the 4LAC sources and vice versa, and we verified that the two cross-matching catalogues contained the same sources.

\subsection{Cross-match 4LAC-AT20G}
\label{subsec:Cross-match_4LAC-AT20G}

From the cross-match between the 4LAC (2682 objects) and the AT20G (5890 objects), we obtained a sample composed of 677 gamma-ray blazars detected at high radio frequency. We determined a matching radius of 7\arcsec\ by using the same method adopted to create the sample 4LAC-GLEAM (see Sect.\ \ref{subsec:Cross-match_4LAC-GLEAM}). Even in this case values above the matching radius chosen exceeds the number of the real matches.

\subsection{Cross-match 4LAC-GLEAM vs.\ 4LAC-AT20G}
\label{subsec:Cross-match 4LAC-GLEAM_vs._4LAC-AT20G}

Eventually, we cross-matched the sample of 1274 blazars detected at low radio frequency (see Sect.\ \ref{subsec:Cross-match_4LAC-GLEAM}), with the sample of 677 blazars detected at high radio frequency (see Sect.\ \ref{subsec:Cross-match_4LAC-AT20G}). This cross-match was realised based on the coincidence of the 4LAC counterpart, therefore it did not require any positional matching (i.e., we assumed that whenever a gamma-ray blazar has both a GLEAM and an AT20G counterpart, the two are associated with each other). This final catalogue is composed of 612 objects.

\begin{table}
\caption{\label{tab:det_rate} Detection rates for 4LAC sources, in MWACS and GLEAM catalogues.}
\centering
\renewcommand{\arraystretch}{1.3}
\begin{tabular}{l c c c c c c}
\hline \hline
      && \multicolumn{2}{c}{3LAC-MWACS} && \multicolumn{2}{c}{4LAC-GLEAM}   \\      
Class && Ratio       & Det.\ rate  && Ratio          & Det.\ rate \\
\hline
total && 87\,/\,247  & 35\%        &&  1274\,/\,1827 & 70\%       \\
FSRQ  && 52\,/\,71   & 73\%        &&  406\,/\,440   & 92\%       \\
BCU   && 16\,/\,89   & 18\%        &&  456\,/\,689   & 66\%       \\
BLL   && 19\,/\,87   & 22\%        &&  384\,/\,659   & 58\%       \\
LSP   && 67\,/\,128  & 52\%        &&  716\,/\,825   & 87\%       \\
ISP   && 11\,/\,37   & 30\%        &&  130\,/\,255   & 51\%       \\
HSP   && 9\,/\,68    & 13\%        &&  115\,/\,248   & 46\%       \\
\hline
\end{tabular}
\end{table}

\section{Results}
\label{sec:Results}

In this section we report the results about the detection rates and the spectral properties of the gamma-ray blazars detected with GLEAM at low radio frequencies, and considerations about emission from the core and the extended region of AGNs. 

\subsection{Detection rate: a step forward from 3LAC-MWACS to 4LAC-GLEAM}
\label{subsec:Detection_rate_a_step_forward_from_3LAC-MWACS_to_4LAC-GLEAM}

In Table \ref{tab:det_rate} we report the values of the detection rates for gamma-ray blazars, in MWACS and GLEAM catalogues. Data for MWACS and 3LAC catalogues are reported from \citetalias{Gir2016}. We summarise these results in the following points:

\begin{enumerate}

  \item The detection rate increased to 70\% for 4LAC-GLEAM compared to 35\% for 3LAC-MWACS. In comparison to the previous study based on the commissioning survey \citepalias{Gir2016}, these constitute an $14.6\times$ and $2\times$ increase in the total number of sources and in the detection rate, respectively. Noteworthy, the detection rate grows for every kind of objects, both considering the SED and spectroscopic classification. Nearly all (92\%) FSRQs are now detected. The improvement in the detection rate is especially impressive for the faintest classes such as the BL Lacs (from 22\% in 3LAC-MWACS to 58\% in 4LAC-GLEAM) and the BCUs (from 18\% in 3LAC-MWACS to 66\% in 4LAC-GLEAM). 
 
  \item Analysing the detection rate of sources classified following the SED, the largest group of detected sources is the one of the LSPs, followed by the ISPs and the HSPs. This is not surprising, as the LSP is the biggest group of sources in the starting sample (4LAC), and its detection rate was already the greatest in the 3LAC-MWACS cross-match.
  
  \item Analysing the detection rate of sources classified following the spectroscopic classification, we notice that the largest detection rate is the one of the FSRQs, while BL Lacs and BCUs have similar detection rates. This confirms the optical spectroscopic campaigns carried out in recent years and already mentioned in Sect.\,\ref{subsec:4LAC_catalogue:_a_subset_of_the_4FGL}  (\citet{Massaro2016opt}, \citet{Marchesi2018}, \citet{Paino2017I}, \citet{Paiano2017II}, \citet{Pena-Herazo2017} and \citet{Alvarez2016}). 

\end{enumerate}

\subsection{Detection rate as function of the radio and gamma emission }
\label{subsec:Detection_rate _as _function _of _the _radio _and_gamma_emission}

\begin{figure}
	\begin{minipage}{0.48\textwidth}
	\hspace{-0.2cm}
		\includegraphics[scale=0.5]{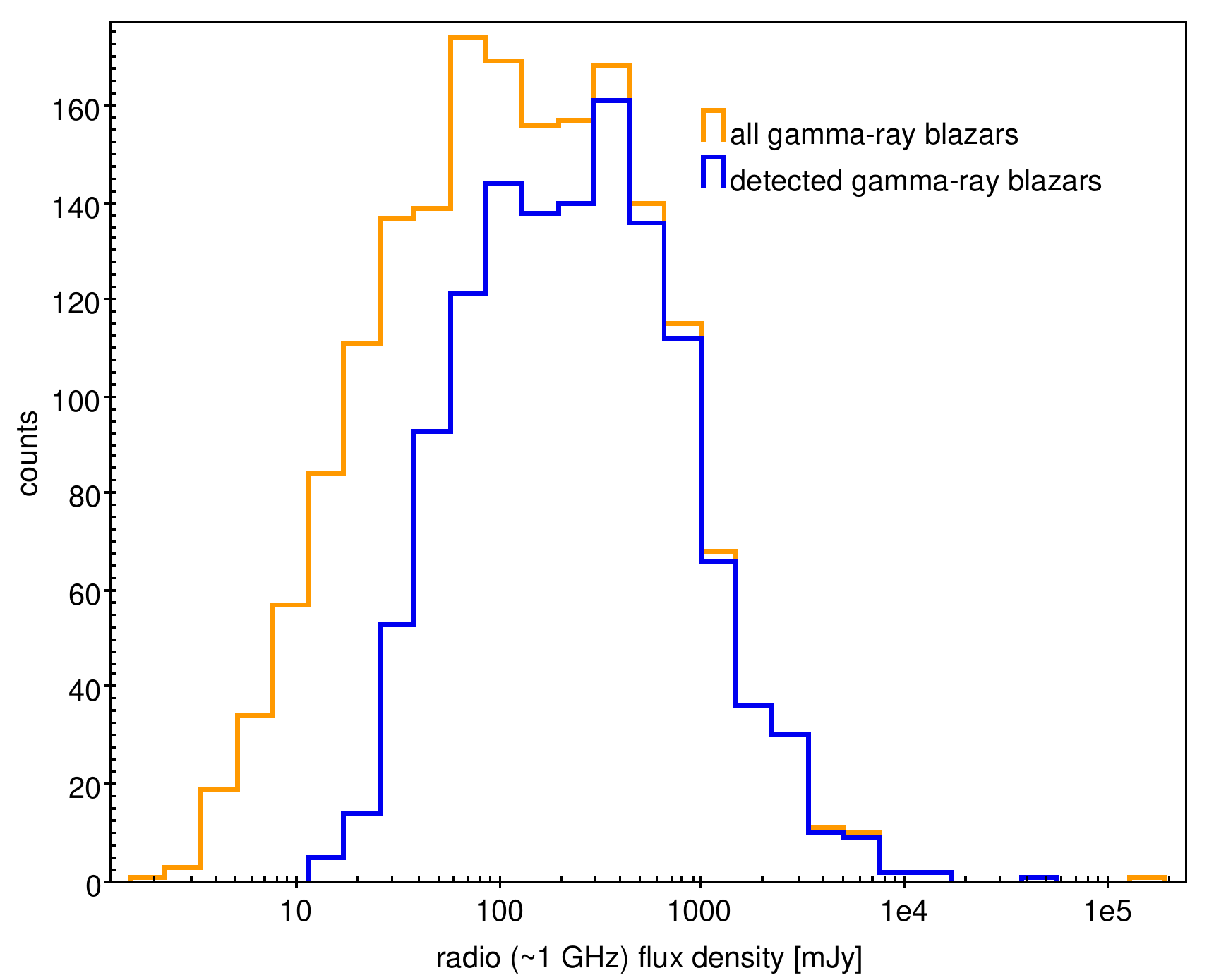}
		\caption{ \small {\label{fig:Comp_1GHz} Distribution of the 1\,GHz flux densities for the entire 4LAC sample in the GLEAM sky (orange) and for the 4LAC sources with GLEAM counterpart (blue). Notice the gamma-ray blazars with flux density larger than the GLEAM sensitivity (6--10\,mJy/beam), that resulted undetected in our cross-matching. These sources must have an inverted spectrum.}}
	\end{minipage}
\end{figure}

\begin{figure}
	\begin{minipage}{0.48\textwidth}
	\hspace{-0.2cm}
		\vspace{+0.1cm}
		\includegraphics[scale=0.497]{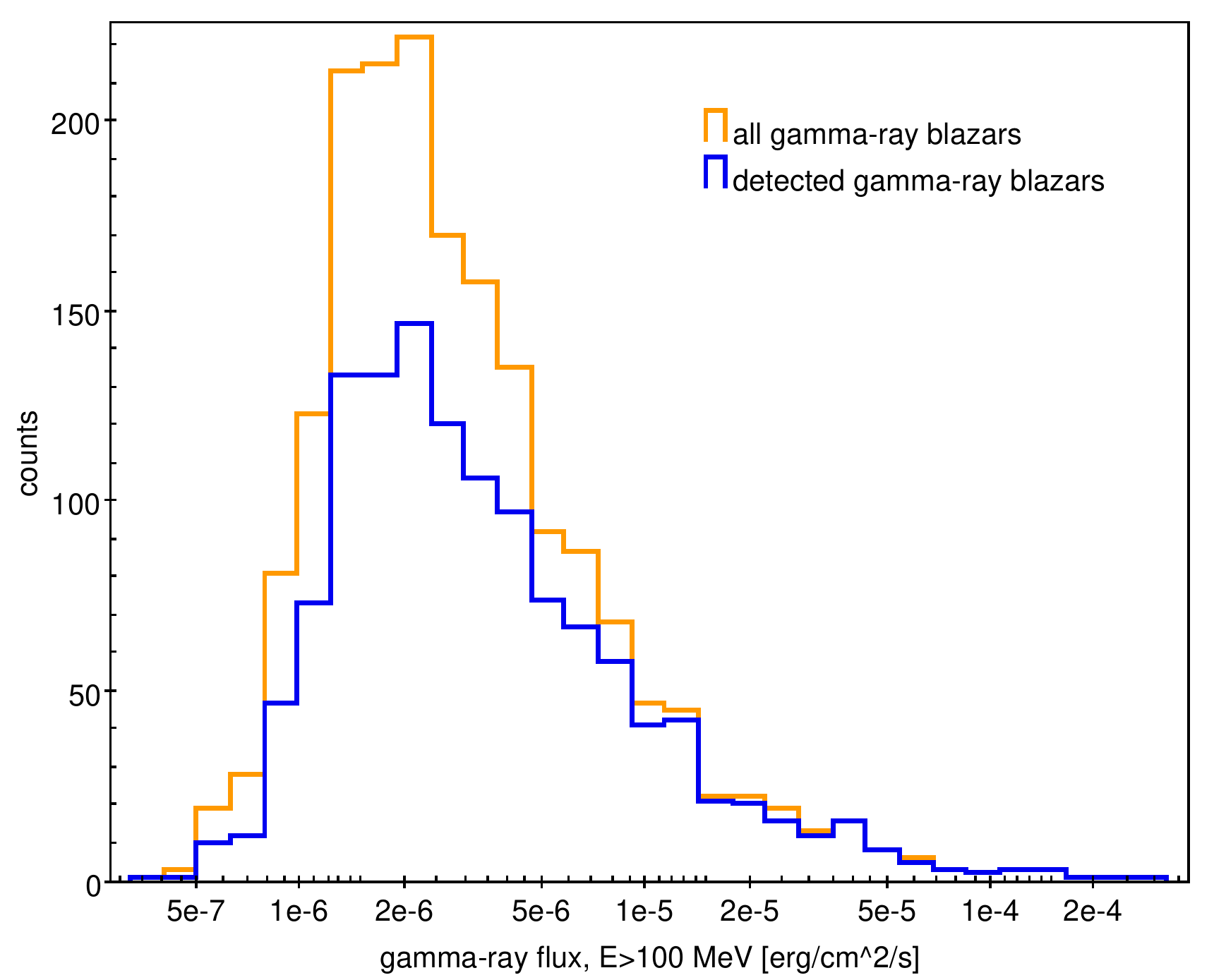}
		\caption{ \small {\label{fig:Comp_gamma} Distribution of the gamma-ray flux (at $E$ > 100\,MeV) for the entire 4LAC sample in the GLEAM sky (orange) and for the 4LAC sources with GLEAM counterpart (blue).}}
	\end{minipage}
\end{figure}

Figures \ref{fig:Comp_1GHz} and \ref{fig:Comp_gamma} show, respectively, the detection rate as function of the flux density at the GHz regime, and as function of the gamma-ray energy flux at $E > 100$\,MeV. In fact, in these histograms we compared the 4LAC sample with the GLEAM sample. We describe here the relative results: 

\begin{enumerate}

  \item  The detection rate grows with the brightness of the objects, both in the radio and in the gamma-ray band. This behaviour was quite predictable. The gamma-ray source in Fig.\ \ref{fig:Comp_1GHz} with a powerful radio flux density ($S_{1\,\mathrm{GHz}} \sim 10^5$\,mJy) that is not detected in GLEAM is the famous radio galaxy Centaurus A, lying at very low redshift \mbox{($z$ = 0.004)} and with bright radio lobes. This source is not included in the GLEAM catalogue because of its bright side-lobes. As radio-galaxies constitute a tiny fraction of the 4LAC which we are not considering in this work, this missing match is not a concern. 

  \item  In Fig. \ref{fig:Comp_gamma} the distribution of the gamma-ray flux densities has the typical shape of the complete source-count distributions: the number of detected sources increases regularly as the flux density decreases, until a suddenly turn over is reached at the instrument sensitivity limit. In the same way, the brightest gamma-ray sources almost always have a GLEAM counterpart, and the GLEAM detection rate decreases only for fainter sources. 
  
  \item In Fig. \ref{fig:Comp_1GHz} the distributions of the radio flux densities is by far less symmetric, and lacks of any well defined turn over at low flux densities.  
  
  \item In Fig. \ref{fig:Comp_1GHz} an interesting feature is the rather large number of gamma-ray blazars with GHz radio flux density larger than the GLEAM sensitivity (6--10\,mJy/beam), that actually resulted undetected in our cross-matching. These sources must have an inverted spectrum.

\end{enumerate}

\subsection{Spectral index}
\label{subsec:Spectral_index}

\begin{figure*}
		\includegraphics[width=\columnwidth]{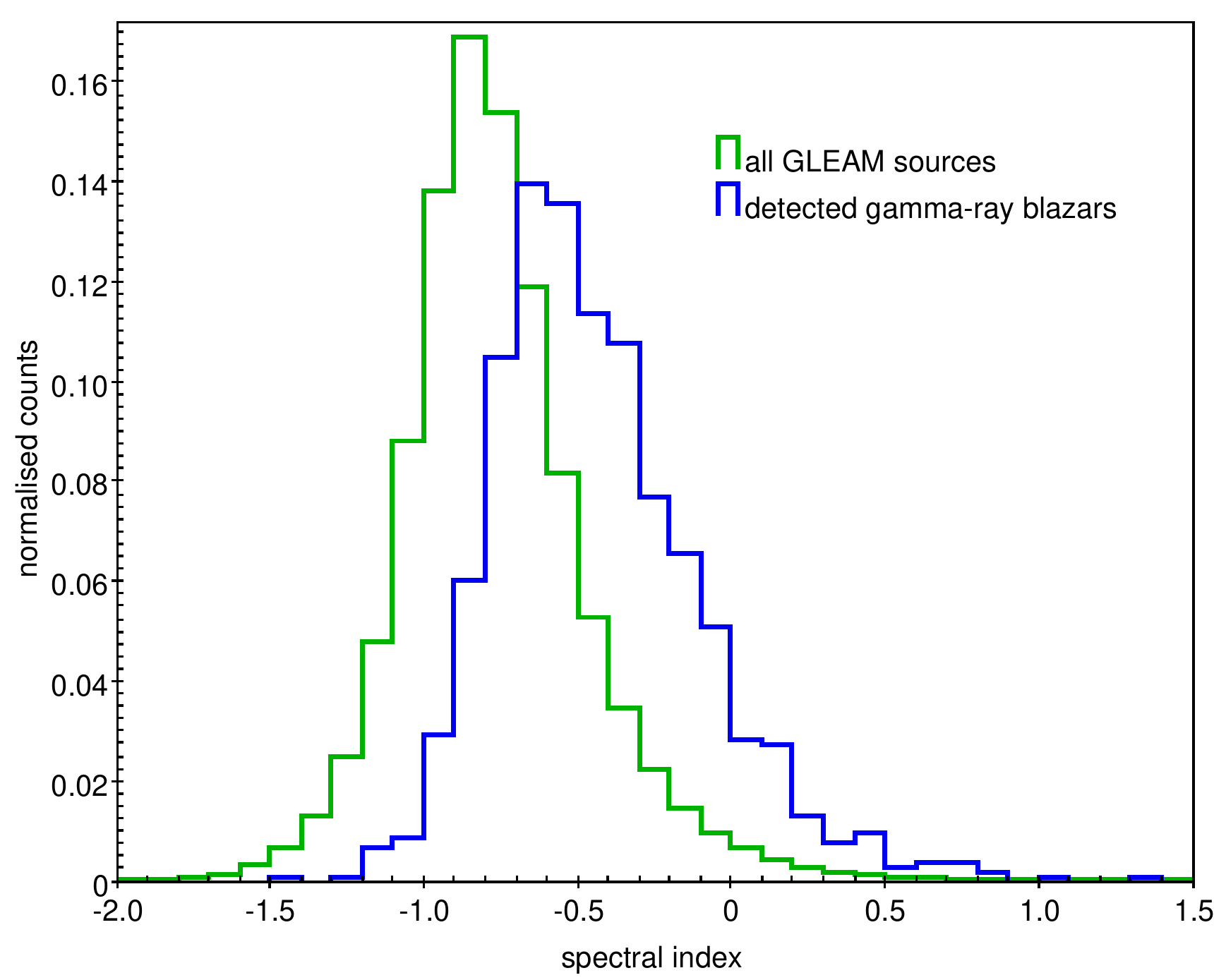}
		\includegraphics[width=\columnwidth]{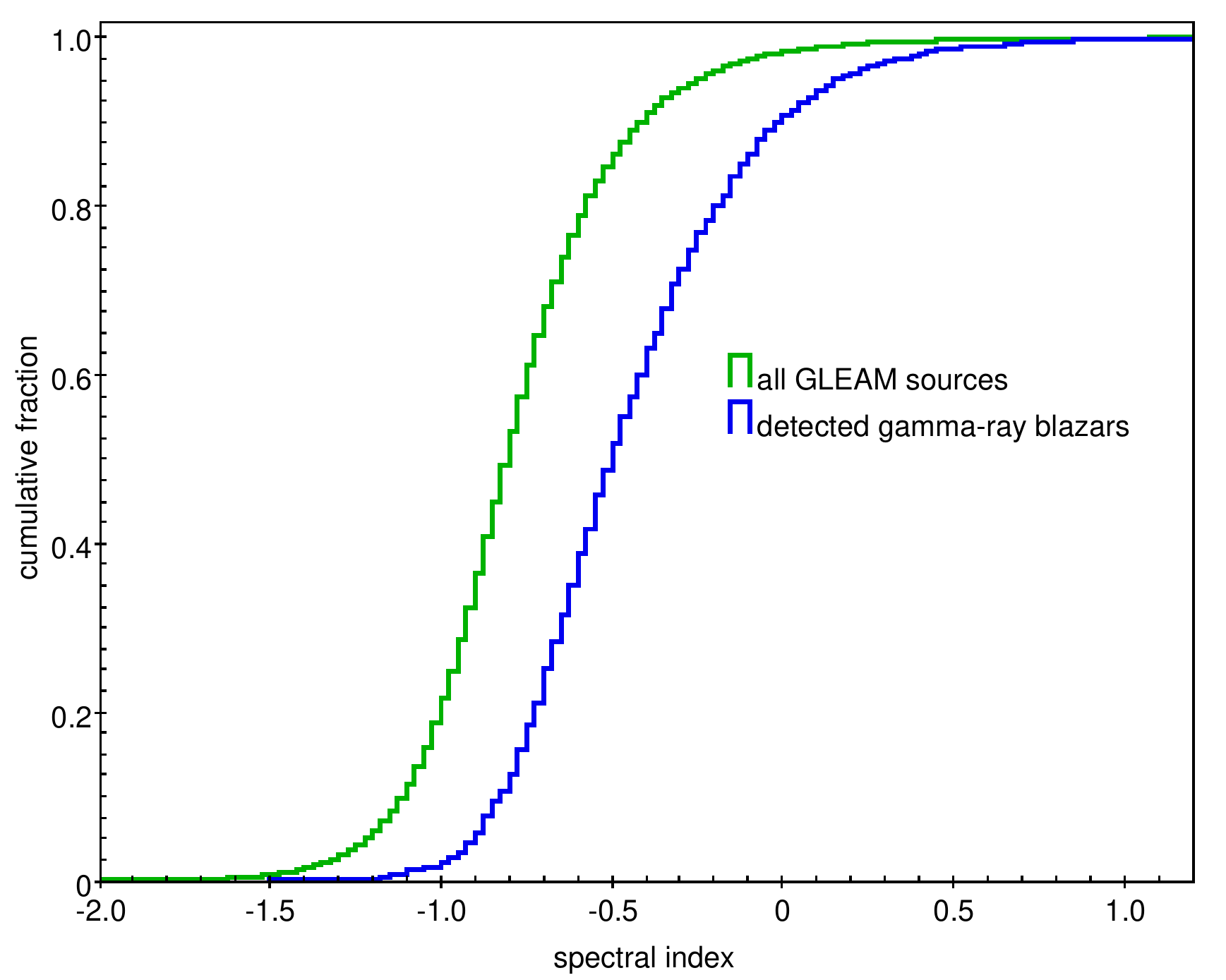}
		\caption{ \small {\label{fig:Comp_SI_norm} Probability density (left) and cumulative distribution (right) functions of the spectral indices ($\alpha$), both for the blazars detected at low frequencies (blue) and for all the GLEAM sources (green). }}
\end{figure*}

The population of gamma-ray blazars detected at low radio frequencies (i.e., in the 72--231\,MHz range), presents a radio spectrum flatter than the rest of the GLEAM population. This result is shown in Fig.\,\ref{fig:Comp_SI_norm}, where are plotted the normalised and the cumulative distributions of the GLEAM blazars spectral indices, both for the whole survey and for the subset of detected sources.
In table \ref{tab:Spectral_Indices_Ranges} we reported the mean spectral indices for different frequency ranges (76-115 MHz,115-151 MHz, 151-189 MHz and 189-227 MHz). Even this table shows that gamma-ray blazars spectra are flatter than all the rest of the GLEAM sources.


\begin{table*}
\caption{\label{tab:Spectral_Indices_Ranges} {Mean values of the spectral indices in the frequency ranges 76--115\,MHz, 115--151\,MHz, 151--189\,MHz, and 189--227\,MHz, for the whole GLEAM population and for the population of detected blazars. For each frequency interval are reported also the total number of sources for which it is possible to estimate the spectral indices. The errors have been computed as standard deviations of the mean}.}
\hspace{-0.5cm}
\renewcommand{\arraystretch}{1.3}
\begin{adjustbox}{max width=1.00\textwidth}
\begin{tabular}{l c c c c c c c c c c c c c}
\hline \hline
\multirow{2}{*}{Population} & \multirow{2}{*}{Sources} & &            \multicolumn{2}{c}{76--115\,MHz}                             & &           \multicolumn{2}{c}{115--151\,MHz}                             & &           \multicolumn{2}{c}{151--189\,MHz}                             & &           \multicolumn{2}{c}{189--227\,MHz}                             \\
                            &                          & & Sources  & $\langle \alpha \rangle \pm \sigma_{\langle \alpha \rangle}$ & & Sources  & $\langle \alpha \rangle \pm \sigma_{\langle \alpha \rangle}$ & & Sources  & $\langle \alpha \rangle \pm \sigma_{\langle \alpha \rangle}$ & & Sources  & $\langle \alpha \rangle \pm \sigma_{\langle \alpha \rangle}$ \\
\hline
GLEAM                       & 307\,455                 & & 281\,817 & $-$0.62 $\pm$ 0.03                                           & & 299\,743 & $-$0.63 $\pm$ 0.04                                           & & 304\,161 & $-$0.52 $\pm$ 0.04                                           & & 304\,637 & $-$0.67 $\pm$ 0.05                                           \\
blazars                     & 1274                     & & 1188     & $-$0.38 $\pm$ 0.04                                           & & 1252     & $-$0.36 $\pm$ 0.04                                           & & 1266     & $-$0.29 $\pm$ 0.04                                           & & 1267     & $-$0.36 $\pm$ 0.05                                           \\

\hline
\end{tabular} 
\end{adjustbox}
\end{table*}

In Table \ref{tab:Spectral_Indices} we report the mean values of the spectral index, for the whole population of the detected blazars, and for the sub-classes obtained based on SED classification and on spectroscopic classification. Notice that the GLEAM catalogue does not report a spectral index for all sources, as some objects have a complex spectra difficult to fit with a single power law. Therefore, for each class we report both the total number of objects, and the number of objects actually provided with a value of the spectral index.

\begin{table}
\caption{\label{tab:Spectral_Indices} Mean values of the spectral index for the whole population of the detected blazars, and for the sub-classes obtained based on SED classification and on spectroscopic classification. We report both the total number of blazars detected at low frequencies in GLEAM, and the number of objects for which GLEAM provides a values for the spectral index $\alpha$. The associated errors have been computed as standard deviations of the mean.}
\centering
\renewcommand{\arraystretch}{1.3}
\begin{tabular}{l c c c}  
\hline \hline
Class & Total   & Sources with   & $\langle \alpha_{\mathrm{GLEAM}} \rangle \pm \sigma_{\langle \alpha_{\mathrm{GLEAM} }\rangle}$ \\
      & sources & spectral index &                    \\
\hline
total &   1274 &     1067       & $-$0.44 $\pm$ 0.01 \\
FSRQ  &   406  &      372       & $-$0.36 $\pm$ 0.02 \\
BCU   &   456  &      367       & $-$0.50 $\pm$ 0.02 \\
BLL   &   384  &      302       & $-$0.44 $\pm$ 0.02 \\
LSP   &   716  &      637       & $-$0.37 $\pm$ 0.01 \\
ISP   &   130  &      94        & $-$0.49 $\pm$ 0.03 \\
HSP   &   115  &      85        & $-$0.56 $\pm$ 0.03 \\
\hline
\end{tabular}
\end{table}

\begin{table}
\caption{\label{tab:frequency_ranges} Observed and emitted frequency ranges for every bin of redshift. }
\centering
\renewcommand{\arraystretch}{1.3}
\begin{tabular}{l c c }  
\hline \hline
Central  & observed  & emitted  \\ 
redshift & frequencies & frequencies                     \\
in bin    &  [MHz]      &   [MHz]                           \\
\hline
1.2      &  99-107     &   218-235   \\
1.0      &  107-115    &   214-230   \\
0.8      &  115-130    &   207-234    \\
0.6      &  130-143    &   208-229    \\
0.4      &  151-166    &   211-232    \\
0.2      &  174-189    &   209-227    \\

\hline
\end{tabular}
\end{table}

\begin{table}
\caption{\label{tab:Intrinsic_Spectral_Indices} Mean values of the intrinsic spectral index, and for the sub-classes obtained based on SED classification and on spectroscopic classification. We report  the number of objects for which we have been able to provide the values for the rest frame spectral index $\alpha_{\mathrm{RF}}$. The associated errors have been computed as standard deviations of the mean.}
\centering
\renewcommand{\arraystretch}{1.3}
\begin{tabular}{l c c c}  
\hline \hline
Class &  Sources at   & $\langle \alpha_{\mathrm{r.f.}} \rangle \pm \sigma_{\alpha_{\mathrm{r.f.}}}$ \\
      &  rest frame &                    \\
\hline
total &      1277  & $-$0.32 $\pm$ 0.03 \\
FSRQ  &      523   & $-$0.29 $\pm$ 0.04 \\
BCU   &      348   & $-$0.38 $\pm$ 0.06 \\
BLL   &      372   & $-$0.29 $\pm$ 0.06 \\
LSP   &      789   & $-$0.31 $\pm$ 0.04 \\
ISP   &      115   & $-$0.21 $\pm$ 0.11\\
HSP   &      101   & $-$0.31 $\pm$ 0.10 \\
\hline
\end{tabular}
\end{table}

\begin{figure}
		\includegraphics[width=0.5\textwidth]{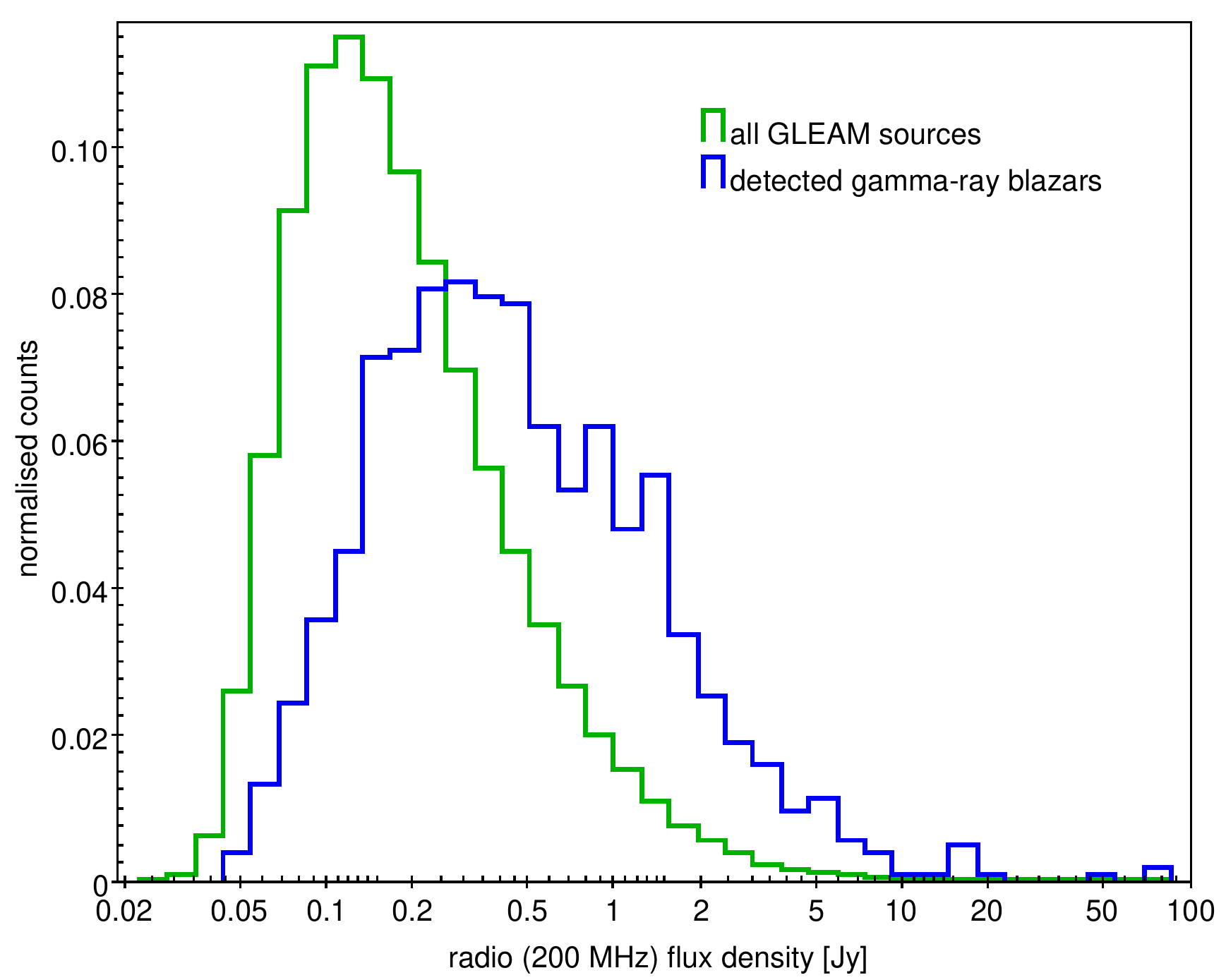}
		\caption{\label{fig:flux_dens_distrib} Low-frequency flux density distribution: blazars are brighter then the rest of GLEAM sources.}
\end{figure}


From Table \ref{tab:Spectral_Indices}, it is possible to see that the spectral indices characteristic of LSPs and HSPs are {\bf significantly} different $\Delta\alpha=\alpha_\mathrm{LSP} - \alpha_\mathrm{HSP}= 0.19 \pm 0.04$. To understand if this gap is an intrinsic difference between the two classes, we first have to remove the possible dependence on the different redshift distributions for the two types (with LSP blazars being typically at higher $z$ than HSP).  We therefore consider the rest-frame spectral indices, taking advantage of the wide spectral coverage that characterises the GLEAM survey, as well as of its spectral resolution that provides 20 separate flux density measurements in the 72--231\,MHz range (see Sect.\ \ref{subsec:GLEAM_catalogue}). We allocated our blazars into six redshift bins according to their redshift, each  \mbox{$\Delta z$ = 0.2} wide and centred in \mbox{$z$ = 0.2, 0.4, 0.6, 0.8, 1.0, and 1.2}, respectively. To compute the individual spectral indices, then, we assumed the formula  $ \alpha= \log (S_{1}/S_{2}) / \log (\nu_{1}/\nu_{2})$  where we used the integrated radio flux densities measured by GLEAM in different frequency bands, according to the redshift bin.

In detail, for sources in the bin centred at \mbox{$z$ = 0.2}, we computed the spectral index based on the radio flux densities measured in the bands centred at 174 and 189\,MHz (i.e., $S_{174\,\mathrm{MHz}}$ and $S_{189\,\mathrm{MHz}}$); similarly, for sources in the bins centred on $z = 0.4, 0.6, 0.8, 1.0, \mbox{and}\, 1.2$, we used the following pairs: $S_{151\,\mathrm{MHz}}-S_{166\,\mathrm{MHz}}$, $S_{130\,\mathrm{MHz}}-S_{143\,\mathrm{MHz}}$, $S_{115\,\mathrm{MHz}}-S_{130\,\mathrm{MHz}}$, $S_{107\,\mathrm{MHz}}-S_{115\,\mathrm{MHz}}$, and $S_{99\,\mathrm{MHz}}-S_{107\,\mathrm{MHz}}$, respectively.

The reason for this procedure is that these bands of observation, when rest-framed for the corresponding redshift (i.e., $\nu_{\mathrm{rest-frame}} = \nu_{\mathrm{obs}}\ (1 + z)$), all refer to similar intervals of emission frequency, in the range between $\sim$215 and $\sim$230\,MHz. This guarantees that all the computed spectral indices refer to the same spectral region. We are therefore sure that we are comparing consistent area of the radio spectrum. Table \ref{tab:frequency_ranges} displays both the bands of observations and the corresponding emitted range frequency with the corresponding bin of redshift. 


\begin{figure*}
		\includegraphics[width=\columnwidth]{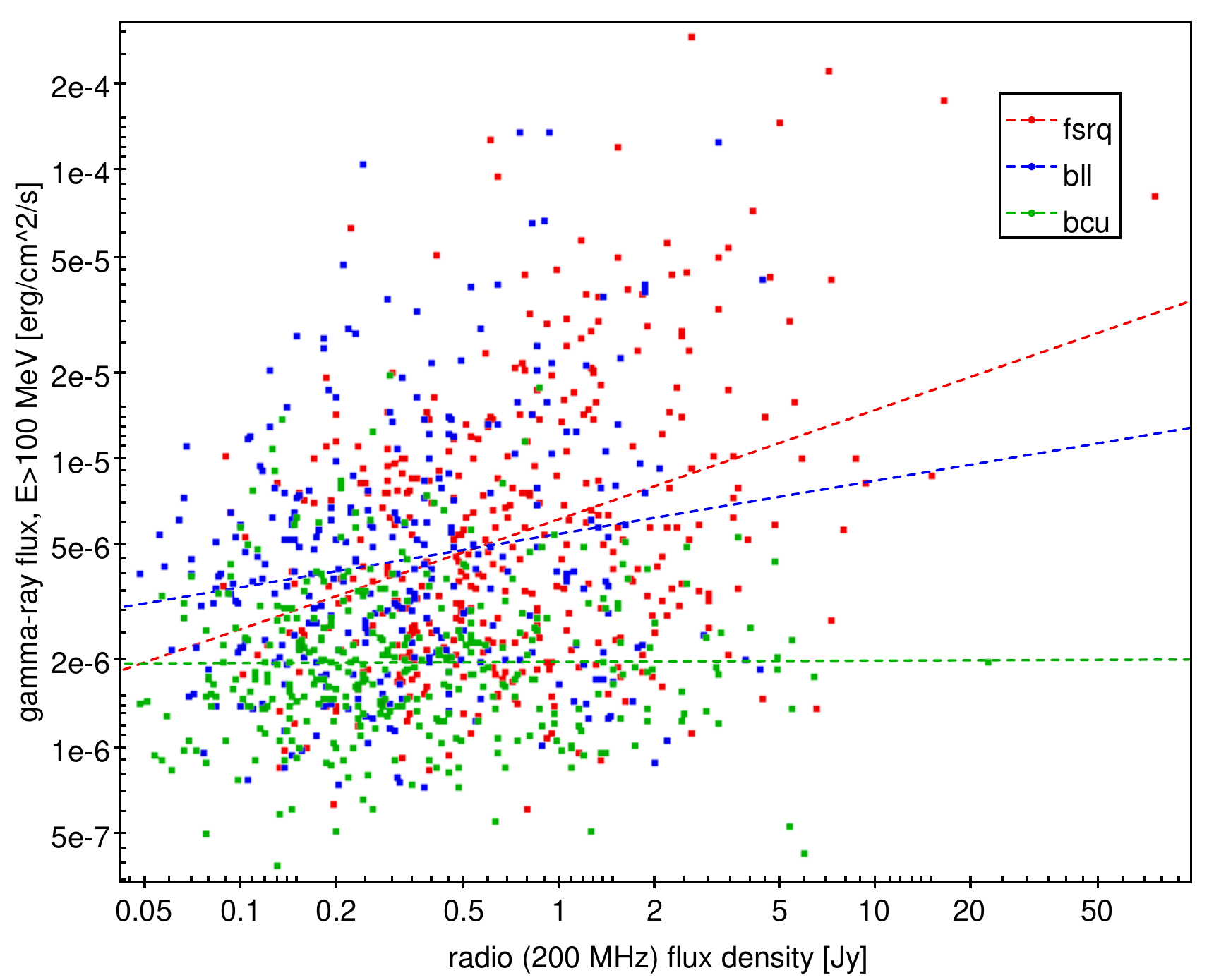}
		\includegraphics[width=\columnwidth]{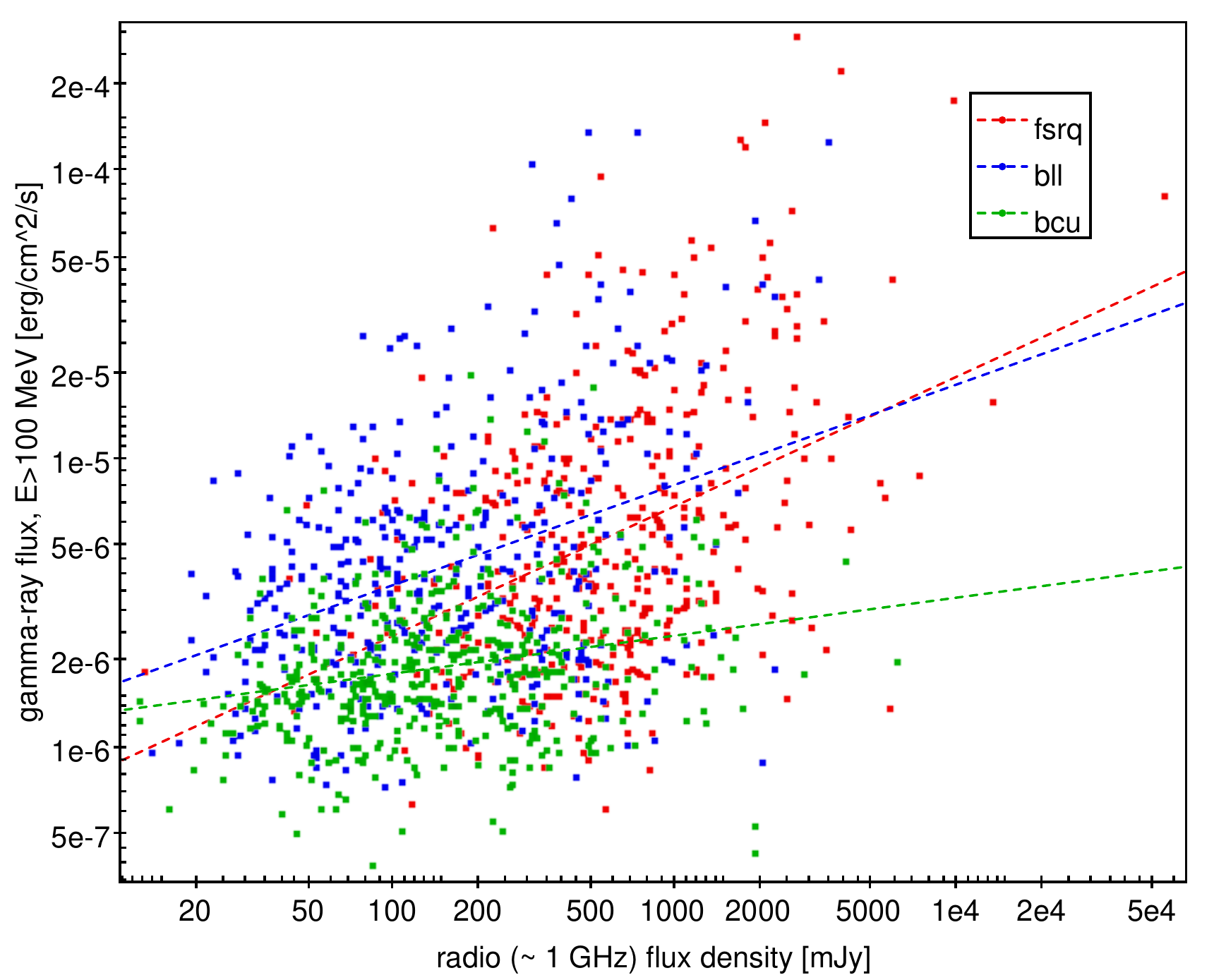}
		\includegraphics[width=\columnwidth]{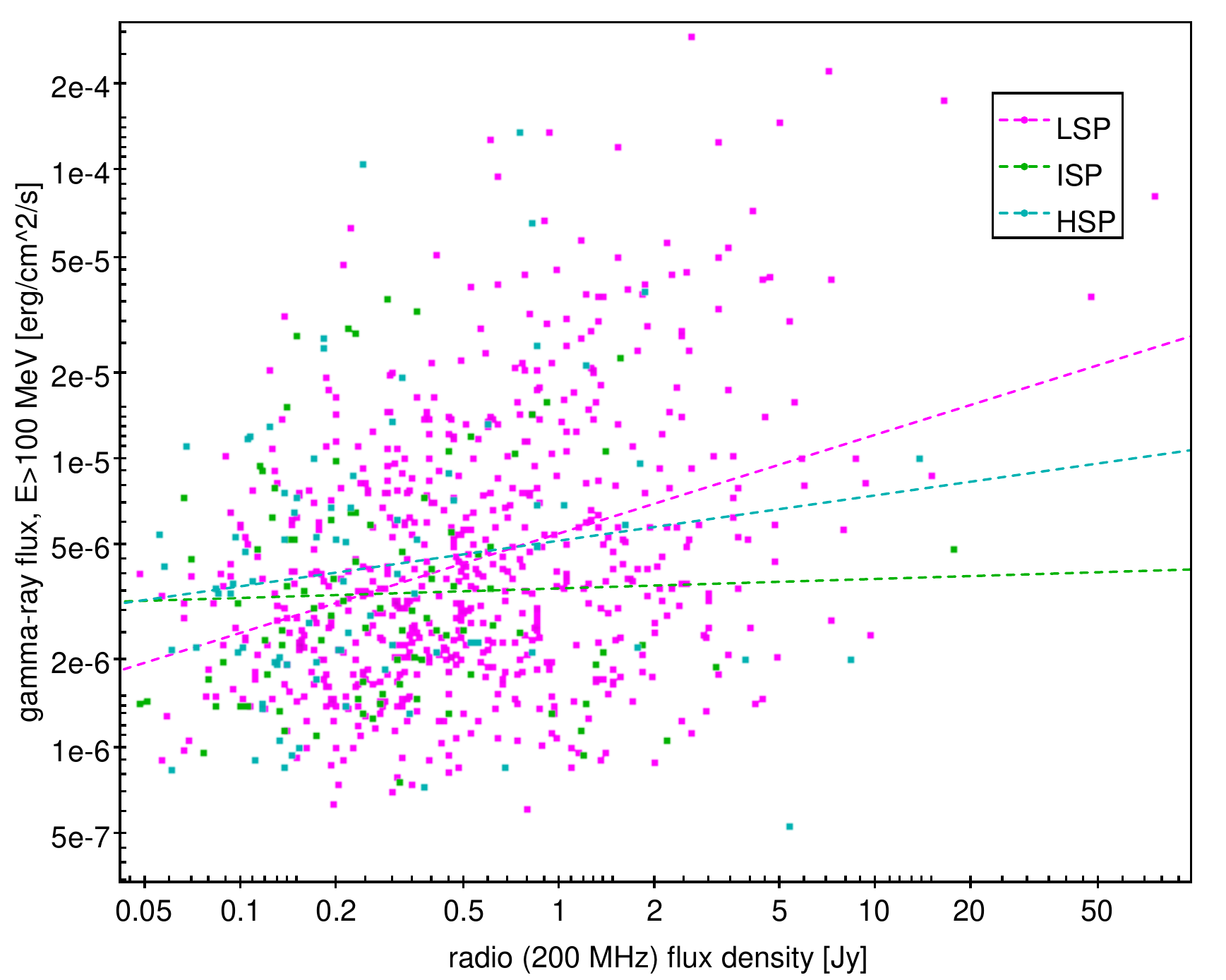}
		\includegraphics[width=\columnwidth]{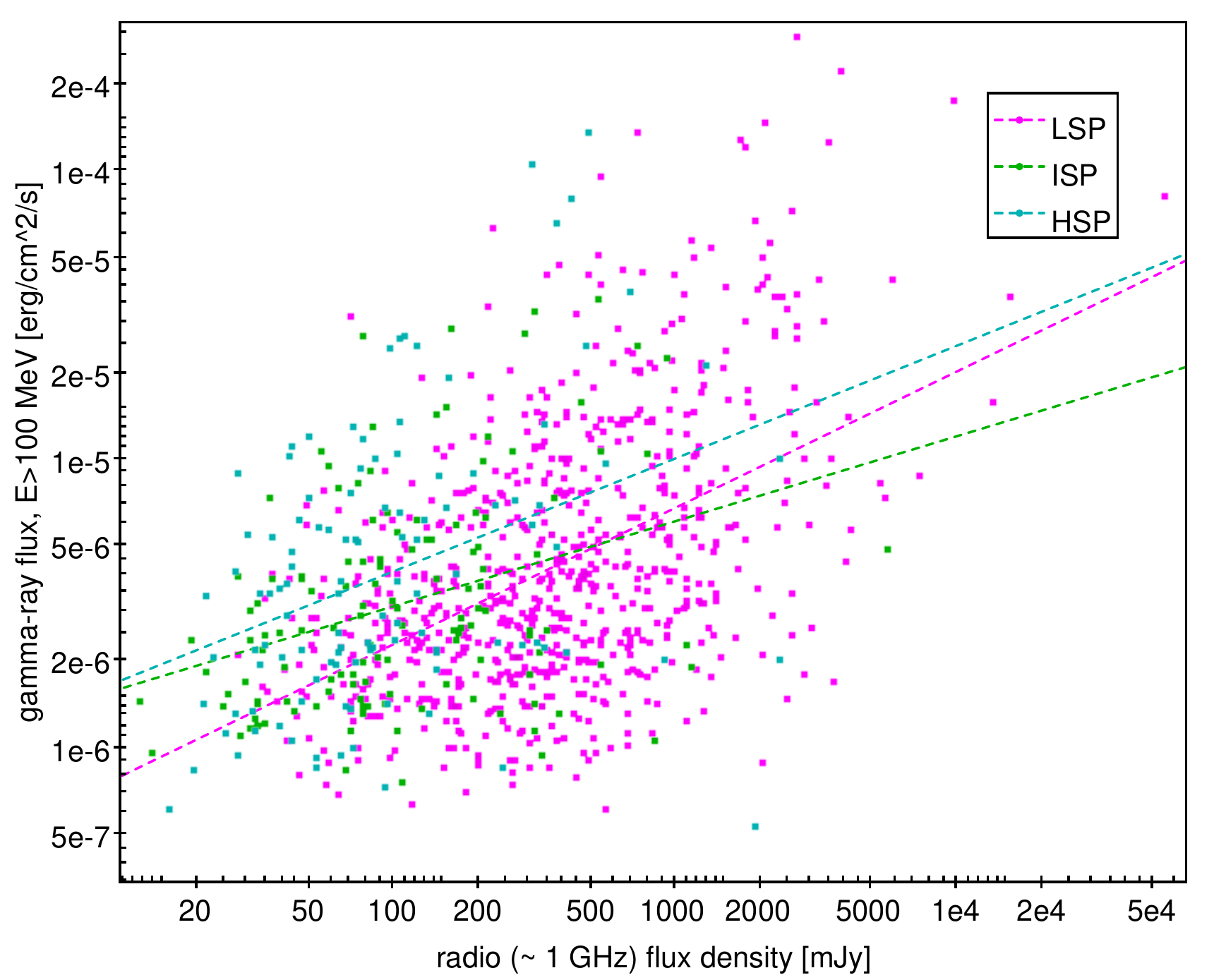}
	\caption{ \small {\label{fig:corr_spectroscopic} Gamma-radio correlations at 200\,MHz (left) and at 1\,GHz (right) for the GLEAM-detected blazars, classified based on their spectra (top panels) or on their SED peak (bottom panels). The radio flux densities at 200\,MHz and at 1\,GHz have been obtained from GLEAM and 4LAC, respectively.}}
\end{figure*}

Table \ref{tab:Intrinsic_Spectral_Indices} finally shows the average values of intrinsic spectral indices for the different classes calculated using the rest frame values in single sub-bands.   The total number of objects in each class has decreased, because we have to discard sources without a reliable spectral index in the relevant sub-band.  Moreover, intrinsic scatter of spectral indexes becomes larger when they are measured in single sub-bands as a result of the poorer signal-to-noise when using narrower frequency intervals.  Since we are narrowing the spectral index to the highest frequency band, all values tend to become flatter.  In any case, the difference between different types are no longer significant, not simply because of the larger standard deviation but also considering the mean value.

\subsection{Gamma-radio correlation}
\label{subsec:Gamma-radio_correlation}

The distribution in Fig.\,\ref{fig:flux_dens_distrib} shows that gamma-ray blazars are brighter in radio, and that they are almost always detected in GLEAM (see Fig.\,\ref{fig:Comp_1GHz}). This finding suggests a very interesting possible correlation between radio and gamma-ray emissions. To test this possibility, we checked if the correlation holds even considering the types of blazars as classified according to their spectra and their SED.

In the top row  of Fig.\ \ref{fig:corr_spectroscopic}, we report the scatter plots of the gamma-ray fluxes as function of radio flux densities at both 200\,MHz (from GLEAM; left Figure) and at $\sim$1\,GHz (from 4LAC; right Figure), with the detected blazars classified according to their optical spectra. In the bottom panels the same quantities are plotted, but in this plot the blazars are classified according to their SED type. The data are distributed with a large scatter; dashed lines report the best fit trend, whose significance needs to be assessed with care.  We have adopted the method presented in \citet{Pavlidou2012}, which takes into account the various limitations involved in dealing with real data. In Table \ref{tab:Corr_coeff} we report the correlation coefficients $r$ and the chances of obtaining a correlation coefficient at least as high as $r$ from uncorrelated data having the same dynamic range in radio flux density, gamma-ray energy flux, and luminosity distance as the observed ones.

From the analysis of the correlation coefficients, it is possible to see that the correlation becomes stronger as the radio frequency increases. This behaviour is in accordance with the results reported in \citet{Ack2011a}, \citet{Mah2010}, \citet{Ghi2010}, \citet{Ghi2011}, and \citetalias{Gir2016}, and is ascribed to the different amount of relativistic Doppler boosting affecting regions with different spectral properties. In detail, the emission at low frequency (200\,MHz) is largely produced in extended lobes, which are not beamed, while there is beaming at higher frequency (1\,GHz) and in the gamma energy flux at $E > 100$ MeV. 



It is noteworthy that the correlation is more important for brighter radio sources, as LSPs and FSRQs. Actually, most FSRQs are also LSPs. However, \citet{Fan2018} finds that the correlation is weaker for FSRQSs than BL Lacs, many of which are HSPs. Evidently, this must be owing to the low depth survey used by the authors, the TGSS, that makes detectable only the brightest BL Lacs. This bias influences the relative core dominance for these sources, showing BL Lacs as more core dominated than FSRQs.

\begin{table}
\caption{\label{tab:Corr_coeff} Coefficients of correlations $r$ at 200\,MHz and 1\,GHz, with the statistical significances $q$-values, estimated using the method of \citet{Pavlidou2012}.} 
\centering
\renewcommand{\arraystretch}{1.3}
\addtolength{\tabcolsep}{+3pt}
\begin{tabular}{l c c c c}
\hline \hline
Class & \multicolumn{2}{c}{200\,MHz}  & \multicolumn{2}{c}{1\,GHz} \\
      &     $r$     &    $q$-value    &     $r$    &    $q$-value  \\
\hline
total &     0.27    & $< 10^{-6} $    &     0.41   & $< 10^{-6}$   \\
FSRQ  &     0.24    & 0.0003 &     0.33   & $< 10^{-6}$   \\
BCU   &     0.03    & 0.88 &     0.16   & 0.55   \\
BLL   &     0.18    & 0.25  &     0.34   & 0.004   \\
LSP   &     0.34    & $< 10^{-6}$     &     0.47   & $< 10^{-6}$ \\
ISP   &     0.12    & 0.59 &    0.29   & 0.14 \\
HSP   &     0.07   & 0.64 &     0.18  &   0.19  \\
\hline
\end{tabular}
\end{table}

\subsection{Core and extended radio emission}
\label{subsec:Core_and_extended_radio_emission}

The total flux density detected from a radio source can be separated into two spectrum components, following the relation:

\begin{equation} \label{eq.2}
\qquad S_\nu= S_{\nu,\ \mathrm{lobe}} + S_{\nu,\ \mathrm{core}} = K_\mathrm{lobe}\,\nu^{\alpha_\mathrm{lobe}} + K_\mathrm{core}\,\nu^{\alpha_\mathrm{core}}
\end{equation}

\noindent where $K_\mathrm{lobe}$ and $K_\mathrm{core}$ are two coefficients that describe the lobe and the core fraction of the total flux density, respectively \citepalias{Gir2016}.

To estimate the emission ratio between the steep spectrum lobe and the flat spectrum core, we made use of the definition of spectral index in the frequency range of GLEAM ($ \alpha= \log (S_{1}/S_{2}) / \log (\nu_{1}/\nu_{2})$\noindent), where $\nu_{1} = 72$\,MHz and $\nu_{2} = 231$\,MHz.

The previous relations allow us to express the ratio between $K_\mathrm{lobe}$ and $K_\mathrm{core}$ as follows:

\begin{equation} \label{eq.3}
\qquad \frac{\,K_\mathrm{lobe}\,}{K_\mathrm{core}} = \frac{\ -10^{\alpha\,\log{(\nu_{1}/\nu_{2})}} \times \nu_{2}^{\alpha_\mathrm{core}} + \nu_{1}^{\alpha_\mathrm{core}}\ } {10^{\alpha\,\log{(\nu_{1}/\nu_{2})}}\times\nu_{2}^{\alpha_\mathrm{lobe}}-\nu_{1}^{\alpha_\mathrm{lobe}}} 
\end{equation}

We estimate $\alpha_\mathrm{lobe}$, $\alpha_\mathrm{core}$, and $\alpha$ based on the mean spectral index of the samples GLEAM ($\alpha_\mathrm{lobe}=-0.77$, which is the value for all sources, which are for the vast majority not-blazars), 4LAC-AT20G ($\alpha_\mathrm{core}=-0.05$) and 4LAC-GLEAM ($\alpha=-0.44$), respectively. 

We estimated a ratio $K_\mathrm{lobe}/K_\mathrm{core} \sim 37$. \citetalias{Gir2016} found that the same ratio was computed with data based on MWACS instead of GLEAM, for a value $\sim$75. The great difference between these two results is due to the limited depth, sensitivity and extension of the MWACS survey, compared with GLEAM. In particular, MWACS can only observe objects with a bright extended emission, overestimating the contribution of these sources. Our present result, on the other hand, shows how the core contribution is more important.

The values of $K_\mathrm{lobe}$ and $K_\mathrm{core}$ by themselves have little intrinsic meaning, as they represent the extrapolation of the relative contribution to zero frequency. However, they are useful to estimate the ratio of the flux density of the core and the lobe emission at each frequency, as provided by Eq.\,\ref{eq.2}. In particular, we determined that at $\sim$158\,MHz, the two spectrum components equally contribute to the total spectrum as $S_\mathrm{lobe}/S_\mathrm{core} \sim 1$.  \citet{Fan2018} found $S_\mathrm{lobe}/S_\mathrm{core} \sim 1$ at $\sim$150\,MHz, which is in good agreement with our result.

\begin{figure}
	\hspace{-0.2cm}
	\includegraphics[scale=0.52]{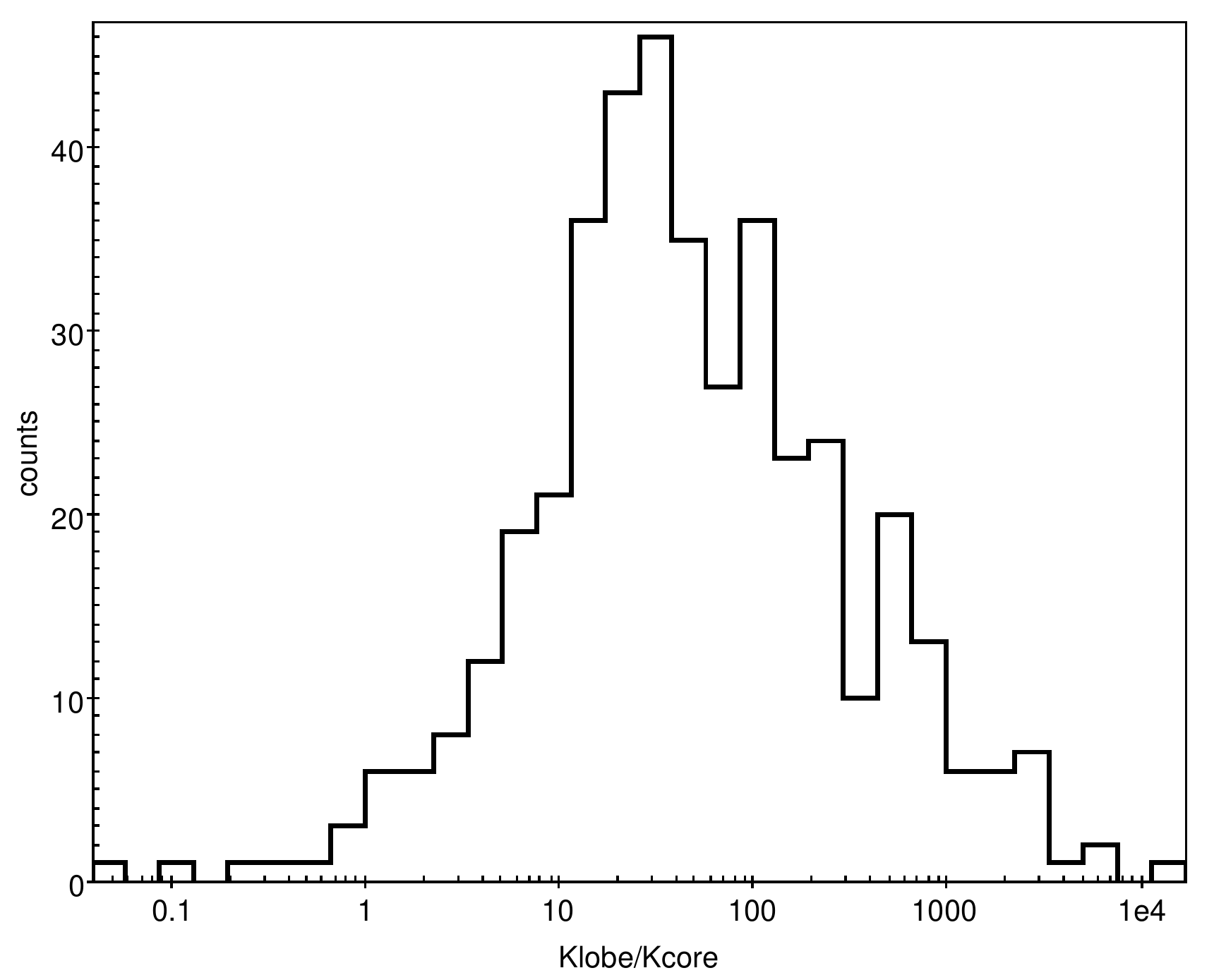}
	\caption{\label{fig:k_dist} Distribution of the ratio $K_\mathrm{lobe}/K_\mathrm{core}$ for the 367 objects of the 4LAC-GLEAM vs.\ 4LAC-AT20G cross-matching.}
\end{figure}

Moreover, we determined the ratio $K_\mathrm{lobe}/K_\mathrm{core}$ for each object of the sample originated cross-matching 4LAC-GLEAM with 4LAC-AT20G, by using the values of the spectral indices for the single source. In Fig.\ \ref{fig:k_dist} we report the distribution of $\log K_\mathrm{lobe}/K_\mathrm{core}$ for all the sources in the sample. The distribution can be described with a Gaussian, with a central value that is in the bin between 26 and 38.

\section{Discussion and conclusions}
\label{sec:Conclusions}

Thanks to the improved sensitivity, bandwidth, and sky coverage that GLEAM provides with respect to MWACS, we have been able to substantially characterise the radio spectrum of gamma-ray blazars in the very low radio frequency band ($\nu <$ 240\,MHz). 

\begin{enumerate}
  \item We have provided more characterised results than the previous paper \citetalias{Gir2016} thanks to the better technical features of the survey GLEAM compared to the survey MWACS. In particular, we have detected more and much fainter sources, which were not detectable by using MWACS due to its limited detection threshold and smaller sky area. We have also increased the number of objects thanks to GLEAM and the 4LAC catalogue. In detail, the total number of gamma-ray blazars detected at low frequency has increased from 87 to 1274 (a factor $14.6\times$) and the detection rate has increased from 35\% to 70\% (a factor $2\times$). FSRQ remains the class with the highest detection rate, reaching as high as 92\%. In other words, almost every gamma-ray FSRQ is detected at low frequency. On the other hand, BL Lacs and BCUs showed a larger fractional increase (from 22\% to 58\% and from 18\% to 66\%, respectively), mainly driven by the fact that BL Lacs tend to be fainter radio sources.

  \item Despite the huge increase in sensitivity (GLEAM is 90\% complete at 170\,mJy, and 50\% complete at 55\,mJy), several blazars that are moderately bright at GHz frequencies (well above the GLEAM completeness levels) remain undetected at $\nu <$ 240\, MHz. Such objects must have inverted spectra or be very variable sources. Another possibility could be that these objects do not have prominent plumes and lobes. In all these scenarios, this indicates that the extended emission contributes very little to their total emission at low frequency.

  \item Blazars are well known for being characterised by flat spectra at high radio frequency, which is actually considered one of their defining features. Our work shows that at low radio frequency these sources continue having flat spectra, both when considered as a population ($\alpha = -0.44\pm0.01$) and when they are divided in the different sub-classes (either when classified based on their SED or their spectra). This result is showed in Fig.\ \ref{fig:Comp_SI_norm}, where we can see that gamma-ray blazars have flatter spectra than the rest of GLEAM sources in the $\sim$100\,MHz regime. This result is not only in agreement with what found by  \citet{Mas2013} and \citet{Nor2014}, but it actually extends those preliminary findings: ($a$) to a much higher statistical significance (as reported in Table \ref{tab:Spectral_Indices}), and ($b$) to an even lower frequency domain. Therefore, we can state that the core contribution to the total radio emission remains substantial, providing also a useful diagnostic for classifying blazars in future sky surveys.

  \item The fact that the core emission still contributes significantly to the total radio flux density also in the GLEAM band, is supported too by the presence of a significant correlation between the gamma-ray energy flux and the low radio frequency flux density. On the other hand, the presence of a much higher scatter in this correlation with respect to those found at higher frequencies \citep{Ack2011a,Mah2010,Ghi2010,Ghi2011} shows that the lobes are also very prominent at low frequency and that they are not significantly correlated with the core emission. In other words, the range of amplification due to the Doppler beaming of the gamma-ray emission is much wider than the intrinsic scatter of the correlation between core and lobe radio power in the rest frame.

  \item The availability of flux density and spectral indexes (both intra- and inter-band) for a sizeable sample of sources at low ($<240$\,MHz, from GLEAM), intermediate ($\sim1$\,GHz, from 4LAC, which in turn got it from NVSS or SUMSS), and high (20\,GHz, from AT20G) frequency, allows us to estimate the relative contributions of the core and lobe radio emission in gamma-ray blazars. We compare the lobe and the core contribution of radio sources by estimating a mean ``zero-frequency'' ratio $K_\mathrm{lobe}/K_\mathrm{core} \sim 37$, which indicates that the core and lobes have about the same power at 158\,MHz. This ratio is lower than the value estimated in \citetalias{Gir2016}. This difference is due to the better technical features of GLEAM, which allowed the detection of sources with flatter spectra (and hence more core-dominated). We can thus predict that even deeper low-frequency surveys will further move the balance to less prominent radio lobes (although for FSRQs we already are near the 100\% detection rate, so we do not expect any dramatic change). Once calibrated with a sizeable sample of sources of different types, the low frequency spectral index could become a quick indicator of the core dominance and therefore a suitable proxy of the Doppler factor for gamma-ray blazars.
      
   \end{enumerate}

\section*{Acknowledgements}

We thank the referee, F. Massaro, for comments and suggestions which were very useful to improve the paper. We thank V.\ Pavlidou for granting us permission to the code used for the computation of the significance of the correlation. We acknowledge financial support through grant PRIN-INAF ob.\ fun.\ 1.05.01.88.06. This research has made use of the SIMBAD database, operated at CDS, Strasbourg, France. This research has made use of NASA's Astrophysics Data System. 

The authors of this work used the catalogues GLEAM and AT20G which were realised thanks to the Murchison Radio-astronomy Observatory and the Australia Telescope Compact Array, over the administration of the CSIRO. The Australian Government provided the necessary financial support for the MWA and the ATCA.




\bibliographystyle{mnras}
\bibliography{Biblio} 





\bsp	
\label{lastpage}
\end{document}